# Shock-induced plasticity of semi-coherent {111} Cu–Ni multilayers

Meizhen Xiang[a,*], Yi Liao[a], Kun Wang[a], Guo Lu[a], Jun Chen[a,*]

[a]Laboratory of Computational Physics, Institute of Applied Physics and Computational Mathematics, Beijing, 100088, CHINA

**Abstract:** Using atomistic simulations, dislocation dynamics modeling, and continuum elastic–plastic stress-wave theory, we present a systematic investigation on shock-induced plasticity in semi-coherent Cu–Ni multilayers. The features of stress-wave evolutions in the multilayers, including wave-front stress attenuation and strong interfacial discontinuities, are revealed by atomistic simulations. Continuum models are proposed to explain the shockwave propagation features. The simulations provide insight into microplasticity behaviors including interactions between lattice and misfit dislocations. The formation of hybrid Lomer–Cottrell locks through the attraction and combination of lattice and misfit dislocations is a major mechanism for trapping gliding lattice dislocations at interfaces. The relationship between dislocation activity and dynamic stress-wave evolution history is explored. The hybrid Lomer–Cottrell locks can dissociate under shock compression or reverse yielding. This dissociation facilitates slip transmission. The influence of coherent stress causes direction dependency in the slip transmission: a lattice dislocation is transmitted more smoothly across an interface from Ni to Cu than from Cu to Ni. The interaction forces between lattice and misfit dislocations are calculated using dislocation dynamics code. Lattice dislocation nucleation from semi-coherent interfaces under shock compression is also reported.

*Principal corresponding author. Tel: (+86)-(010)-61935171.
 *Email addresses:* xiang_meizhen@iapcm.ac.cn (Meizhen Xiang), liaoyi19830512@163.com (Yi Liao), kwang-hnu@163.com (Kun Wang), lu_guo@iapcm.ac.cn (Guo Lu), jun_chen@iapcm.ac.cn (Jun Chen)

# 1. Introduction

Nanoscale metallic multilayers have attracted great attention from both academia and industry. The abundance of interfaces in metallic multilayers yields unique mechanical properties, such as high strength and hardness, improved radiation damage resistance, good thermal stability, increased ductility, and improved fracture toughness (Misra et al. (1998); Clemens (1999); Zheng et al. (2013); Demkowicz et al. (2008a); Zhang et al. (2011); Liu et al. (2011b); Yang and Wang (2016)). Nanoscale metallic multilayers can be fabricated via bottom-up thermodynamic techniques such as physical vapor deposition (Mara et al. (2008)) and the electrodeposition of two different metals (Bakonyi and Peter (2010); Yahalom et al. (1989)). In recent years, many have studied the grain morphologies, interfacial structures, thermal stability, plastic deformation, and failure mechanisms of metallic multilayers (Bauer and Jh (1986); Wan et al. (2012); Misra et al. (2004); Wang et al. (2009); Liu et al. (2013); Wang et al. (2008a,b); Li et al. (2012a); Abdolrahim et al. (2014); Wang and Misra (2014); Kang et al. (2007); Hansen et al. (2013)). These works were recently reviewed by Zhou et al. (2015).

Studies of the shock responses of metallic multilayered materials may be motivated by the potential applicability of such materials in shock-damping devices to attenuate the effects of impacts and blasts. Large differences in the mechanical properties of the component layers induce periodic heterogeneities in multilayer composites that lead to impedance mismatches at the interfaces (Holmes and Tsou (1972); Johnson et al. (1994); Herbold et al. (2008); Neel and Thadhani (2011); Specht et al. (2012); Chiu et al. (2013)). The interaction of shockwaves with these interfaces can change the propagating wave structure and the attenuation of the shockwaves. For this reason, metallic multilayered metals could be applied in structural components that are subjected to strong impact or blast loading, such as high-quality armor and space debris shields in satellites.

The shock responses of composites are generally studied from a macroscopic perspective using experiments and continuum elastic–plastic wave theory. Earlier studies focused on the Hugoniot curves of composites. Using plate impact experiments, Holmes and Tsou studied the Hugoniot curves of unidirectional fiber-reinforced composites (Holmes and Tsou (1972)). Both the shock-wave velocity and the free-surface velocity were measured by optical techniques; the shock front in the composite was found to be steady. Garg and Kirsch utilized concepts from the theory of interacting continua to derive a generalized set of conservation equations (i.e., the Hugoniot relations) across a steadily propagating disturbance in composites (Garg and Kirsch (1971)). Herbold et al. investigated the shock response of polytetrafluoroethylene–Al–W granular composites by experiments and numerical simulations (Herbold et al. (2008)). They found that shock loading of this granular composite induced redistribution of internal energy between components. Such energy redistribution can be tailored by manipulating the particle sizes of the rigid components; higher fractions of thermal energy are transferred to soft and light components as the heavy and rigid components are increased in size. Neel and Thadhani (2011) studied the shock-compression Hugoniot curves of several tetrafluoroethylene–hexafluoropropylene–vinylidene fluoride (THV)–alumina particle composites with varying porosities through plate-impact experiments. The experimental results were correlated to the model predictions by adjusting the "effective" Grüneisen parameter. Kelly and Thadhani (2016) later studied the shock-compression response of highly reactive Ni–Al multilayered thin foils using laser-accelerated thin-foil plate-impact experiments. They also performed mesoscale shock simulations in the CTH programming environment using real imported microstructures of the cross-sections of multilayered Ni–Al foils to compute and correlate the Hugoniot response with the experimentally determined equation of state. At particle velocities below 150 m/s, the experimentally determined equation of state matched the CTH-predicted inert response and showed consistency with the observed unreacted state of the Ni–Al target foils recovered from this velocity regime. At higher particle velocities, the experimentally determined equation of state deviated from the CTH-predicted inert response. Chiu et al. (2013) conducted thick-walled cylinder experiments and simulations of the dynamic collapse of Ni–Al concentric laminate cylinders. They demonstrated the phenomenon of cooperative buckling originating in the innermost layers. The instability of all layers was dictated by the buckling mode of the inner layers. Specht et al. (2012) investigated the wave propagation response of cold-rolled Ni and Al multilayered laminated composites with various orientations by using finite-volume simulations. They found that the layer orientation of dissimilar materials in the multilayered laminated composite greatly affected the dissipation and dispersion of shockwaves. Specht et al. (2017) also measured the Hugoniot curves of NiAl multilayers by plate-on-plate impact experiments. The experimental Hugoniot curve agreed with the finite-volume simulations. Han et al. (2011) studied the shock-induced plasticity behaviors of Cu–Nb multilayers. Comparative study of the shock-induced plastic deformation of the pure Cu and Nb versus that of the Cu–Nb nanolaminates emphasized the importance of the heterogeneous phase interfaces in determining the dynamic deformation behaviors of multilayer materials. Han et al. (2014) also studied the deformation and failure of bulk Cu–Nb nanocomposites under planar shock loading. They found that voids generally nucleated within the Cu phase, which has a higher impedance and lower spall strength than Nb, rather than along the Cu–Nb interfaces or in the Nb phase. This finding contradicted the general theory of failure initiation at interfaces, and indicated that the Cu–Nb interfaces were stable under dynamic loading. Germann et al. (2009) used laser-launched flyer plate experiments and large-scale molecular dynamics (MD) simulations to study the shock responses of Cu–Nb nanoscale multilayered (nanolayered) composites. They observed a strengthening effect at the interfaces under dynamic shock loading, in both MD simulations and post-mortem examinations of shock-recovered samples subjected to 20-GPa shock loading. The MD simulations provided insight into the dislocation nucleation and transmission processes occurring under compression, as well as the subsequent annihilation of said dislocations upon release. Molinari and Ravichandran (2006) analyzed the steady plastic shocks generated by planar impact on metal–polymer laminate composites in the framework of gradient plasticity theory. The following experimental features were well reproduced by gradient plasticity modeling: shock width was proportional to the cell size; the strain rate magnitude was inversely proportional to cell size; and strain rate increase followed a power-law relationship with the applied stress amplitude. While these results were equally well described by first- and second-order gradient plasticity theories, the first-order gradient plasticity approach was favored in comparing the theoretically predicted structure of the shock front to the experimental data.

Studying details of the thermomechanical behaviors of metallic multilayers under shock loading is experimentally challenging, because the reflection and transmission of shockwaves at metallic multilayer interfaces create very complex thermodynamic paths. A complete temporal record of stress, velocity, and temperature is necessary to properly characterize the material state during shock loading. However, such kinetic behavioral information is not readily obtained from current shock experiments. In addition, complex processes triggered by the interactions between shockwaves and interfaces cannot be observed directly or easily by current experimental techniques. MD simulations are an important alternative approach for studying the dynamic behaviors of materials under shock loading (Xiang et al. (2013a,b, 2017)). MD simulations can provide details of atomic-scale mechanisms, which cannot be directly observed in experiments.

Much recent MD simulation-based research has addressed the strengthening effects of interfaces and the underlying micromechanisms (Shao and Medyanik (2010); Shao et al. (2015); Mitchell (2002); Rao and Hazzledine (2000); Wang et al. (2008b,a); Ghosh and Arroyo (2013); Peng and Wei (2016); Zhu et al. (2015); Salehinia et al. (2014); Li et al. (2012b); Zbib et al. (2011); Wang et al. (2014)). However, studies using MD simulations to understand the atomic-scale details of interactions between shockwaves and interfaces have been infrequent. Recently, Zhang et al. (2013) utilized non-equilibrium MD simulations to reveal the dislocation processes underlying the effects of bimetal interfacial structures on the plastic responses of Cu–Nb nanolayered composites to shock compression. Critical shock pressures for the nucleation and transmission of dislocations across atomically flat interfaces were shown to be substantially higher than those from faceted interfaces. In the present work, using {111} Cu–Ni multilayers as an example, we provide atomic-scale details of the shock responses of metallic multilayers. Both the shockwave–interface interactions and the dislocation–interface interactions are investigated in depth by combining atomistic simulations, dislocation modeling, and continuum elastic–plastic wave theory.

As a special class of metallic multilayer materials, the mechanical behaviors of Cu–Ni multilayers have attracted significant recent interest (Misra et al. (1998); Wang et al. (2008b); Liu et al. (2011a); Rao and Hazzledine (2000); Mitchell (2002); Yan et al. (2013); Liu et al. (2012); Yahalom et al. (1989)). Shao et al. (2013, 2014, 2015) recently reported observations from MD simulations of misfit dislocation structures in Cu(111)/Ni(111) semi-coherent interfaces. They found that the interfaces contained several structures, including normal FCC stacking structures, intrinsic stacking-fault structures, misfit dislocations, and misfit dislocation intersections (nodes). The mechanical behaviors of these semi-coherent interfaces under biaxial tension and compression applied parallel to the interfaces were studied by Shao et al. (2015). The nucleation of lattice dislocations occurred preferentially at the nodes, correlated with reduction of excess volume at these points.

The loading method in the present work mimics plate-impact experiments with shock compression applied perpendicular to the interfaces. This differs from the situation in Shao et al. (2015), where loading strains were parallel to the interfaces. The simulations show reflection-loading and -unloading characteristics when the shockwave propagates from Cu to Ni and from Ni to Cu, respectively. This reveals strong interfacial discontinuities in the stress components. By continuum mechanics, these features of the shockwave structures are correlated to mismatches of the Hugoniot curves and the interfacial elastic–plastic properties. The simulations also provide new insight into the behavior of semi-coherent interfaces as barriers for single-dislocation transmission across interfaces. The formation of hybrid stair-rods is identified as a major mechanism barring dislocation transmission across semi-coherent interfaces. In addition, the dissociation of hybrid Lomer–Cottrell locks is found to facilitate the slip transmission of lattice dislocations across semi-coherent interfaces. The influences of coherent stress induce directional dependency in slip transmission: lattice dislocations are transmitted more smoothly across interfaces from Ni to Cu than from Cu to Ni. Dislocation configurations obtained by atomistic simulations are extracted and incorporated into dislocation dynamics simulations, permitting quantitative investigation of the interaction forces between lattice and misfit dislocations. Finally, lattice dislocation nucleation from semi-coherent interfaces under shock compression is reported.

In the following, the simulation details and analysis methods are elaborated in Section 2. The characteristics of the shockwave profiles, including the reflection and transmission of shockwaves at interfaces and interfacial stress component discontinuities, are discussed in Section 3, correlating the non-equilibrium MD (NEMD) simulations to continuum elastic–plastic mechanics. Interactions between gliding lattice dislocations and misfit dislocations are presented in detail in Section 4. Finally, conclusions are drawn in Section 5.

## 2. Model and methodology

Multilayer target samples are constructed by alternately arranging Cu layers and Ni layers along the [111] crystallographic direction. The Cu and Ni layers are set to equal thicknesses. The crystallographic orientations in the Cu and Ni layers are consistent. The $x$, $y$, and $z$ axes of the Cartesian coordinate system align with the [111], [1-10], and [11-2] crystallographic directions, respectively, as shown in Fig. 1(a). The initial separation $h$ between neighboring Cu and Ni layers is set to $h = (a_{Cu} + a_{Ni})/2$, where $a_{Cu} = 3.615$ Å $\times \sqrt{3}/3$ and $a_{Ni} = 3.52$ Å $\times \sqrt{3}/3$ are the lattice constants of Cu and Ni in the [111] direction. The simulation setup consists of an impacting flyer (single-crystal Cu) and a multilayer target, as illustrated in Fig. 1(a). The Cu layers, Ni layers, and interfaces in the multilayer target are labeled as $Cu_1$, $Cu_2$, $Cu_3$, ⋯; $Ni_1$, $Ni_2$, $Ni_3$, ⋯; and $I_1$, $I_2$, $I_3$, ⋯; respectively.

Two specific simulation geometries are designed. The first simulation sample (S1) consists of the impact of a 60-nm-long flyer with a multilayer sample with length $L_x = 120$ nm, width $L_y = L_z = 30$ nm, and layer thickness $W = 30$ nm. To study the stress attenuation effects on larger spatial scales, the second multilayer simulation sample (S2) is designed with the extended length of $L_x = 2400$ nm. The length of the multilayer target in S2 is 20 times larger than that in S1. The layer thickness in S2 is set to 10 nm, such that the multilayer target contains 240 total interfaces.

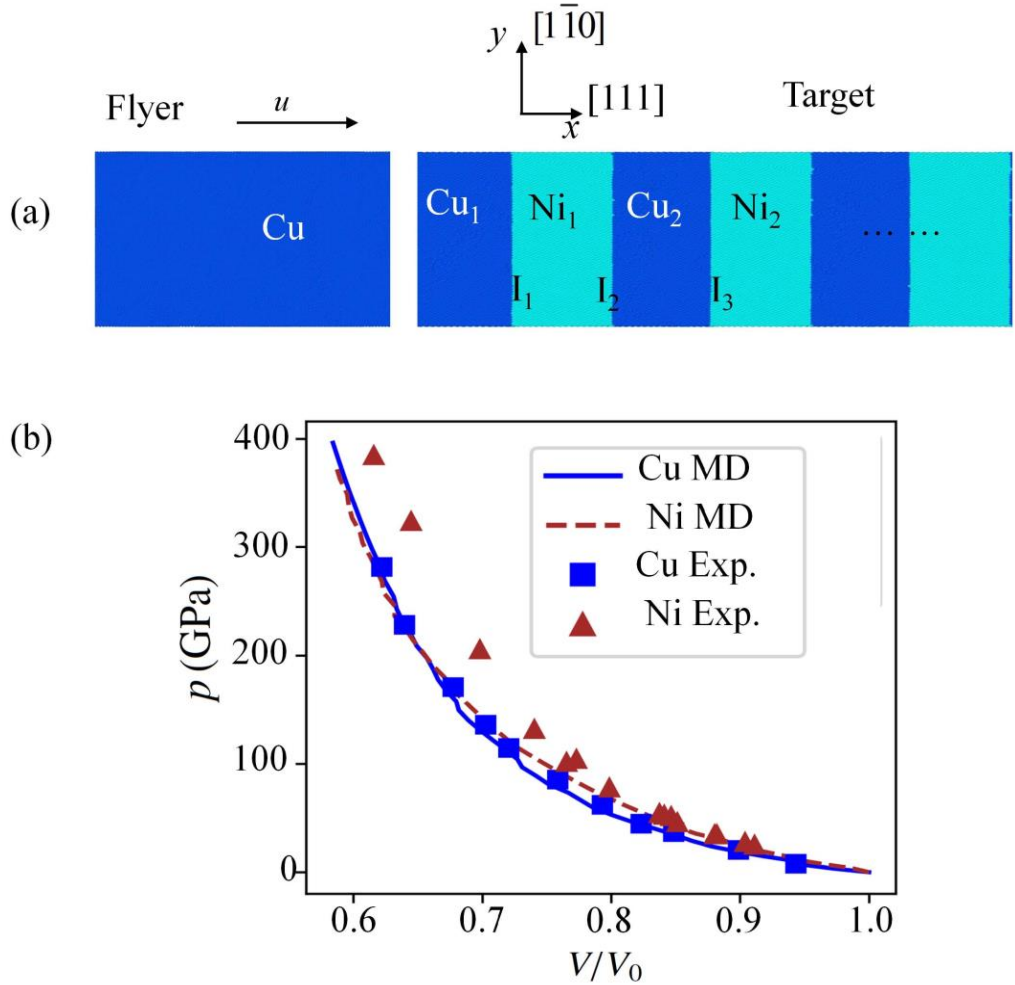

Fig. 1. (a) Illustration of the simulation setup. (b) Comparisons of Hugoniot curves between MD simulations and experimental measurements. The experimental Cu and Ni Hugoniot points are obtained from Marsh (1980) and Demaske et al. (2012), respectively.

The atomic interactions in Cu and Ni are described by an embedded-atom method (EAM) potential developed by Bonny et al. (2009). This potential has been used to simulate the mechanical behaviors of Cu–Ni multilayers under indentation loading (Zhu et al. (2015)). The accuracy of the EAM potential under shock conditions is confirmed by comparing the shock Hugoniot curves calculated by MD simulations with experimental measurements, as shown in Fig. 1(b), where $p$ is pressure and $V/V_0$ is the compression ratio. The calculated Hugoniot curve of Cu agrees perfectly with the experimental measurements. For Ni, the calculated Hugoniot curve coincides well with experimental points for compression ratios $V/V_0 > 0.75$, but underestimates the shock pressure for $V/V_0 < 0.75$ with the relative error of <10%. In the NEMD simulations performed here, $V/V_0 > 0.8$. In this regime, the MD Hugoniot curves of both Cu and Ni agree well with the experimental results.

Before impact simulation, the multilayer targets are relaxed by energy minimization using the conjugate gradient method until the maximum force acting on any atom in the system is ≤5 pN. Then the target samples and the flyer sample are equilibrated using a constant-temperature and -pressure (NPT) ensemble at 300 K and 0 GPa for 50 ps. Fig. 2 shows the semi-coherent interface structures before and after energy minimization and equilibration. The lattice mismatch of Cu and Ni at the interface induces the formation of straight misfit dislocations (green lines in Fig. 2(a,c)) at the interfaces in the un-relaxed sample, as shown in Fig. 2(a). In Fig. 2(a), the misfit dislocations are 1/6 <211> {111}-type Shockley partial dislocations. Similar to the situation in pure face-centered cubic (FCC) metals, the misfit dislocations separate the interface into FCC regions (coherent regions) and intrinsic stacking-fault regions (ISF), marked by hexagonally close-packed (HCP) atoms. Each FCC region or ISF region bounded by the straight misfit dislocations is an equilateral triangle with the side length of ~9.5 nm as quantitatively determined by the lattice mismatch (Shao et al. (2014)). The intersection of the dislocations forms periodically distributed misfit dislocation nodes. The atoms near the node regions have the highest potential energy, as shown in the right-hand image in Fig. 2(a). The FCC region has the lowest energy; the energy of the ISF regions is higher than that of the FCC region, but lower than that of the node region. We assume that the stacking sequence of (111) close-packed planes is $\cdots ABCABC \cdots$ in both the Cu and Ni single-crystal layers. Then, in the FCC, ISF, and central node regions, the three-layer stacking sequences are $A_{Cu}B_{Cu}C_{Ni}$, $A_{Cu}B_{Cu}A_{Ni}$, and $A_{Cu}B_{Cu}B_{Ni}$, respectively, as illustrated in Fig. 2(b). After energy minimization and thermal equilibration, the straight dislocation lines become curved and the interfacial potential energy distribution becomes much more homogeneous than that in the unrelaxed sample, as shown in the right-hand figure in Fig. 2(c).

In addition to the misfit dislocations, some residual lattice dislocations are present in the relaxed sample, as shown in Fig. 2(d). These residual dislocations in the multilayers are generated during the energy minimization process. From Fig. 2(d), the lattice dislocation distribution is not homogeneous in the sample. Instead, dislocations mainly occupy the regions near free surfaces and Cu–Ni interfaces. Furthermore, the amount of residual lattice dislocations in the Cu layers is much larger than that in the Ni layers. If the materials are approximated as isotropic materials, according to Frank's rule, the energy per unit length of a dislocation line is proportional to $\mu b^2$, with $\mu$ being the shear modulus and $b$ being the Burgers vector length. The Burgers vector lengths of partial dislocations in Cu and Ni are subtly different at 1.48 Å and 1.44 Å, respectively. However, the shear modulus of Ni (94.7 GPa) is much higher than that of Cu (54.6 GPa) (Hirth and Lothe (1982)). Thus, the dislocation energy is higher in Ni than in Cu. As a result, it is energetically more feasible for residual lattice dislocations to nucleate in the Cu layers instead of Ni layers during the relaxation processes. In the S1 sample, only two residual lattice dislocations are present in each Ni layer. In the shock simulations, these two residual dislocations in the Ni layer are annihilated as soon as the shockwave front reaches them.

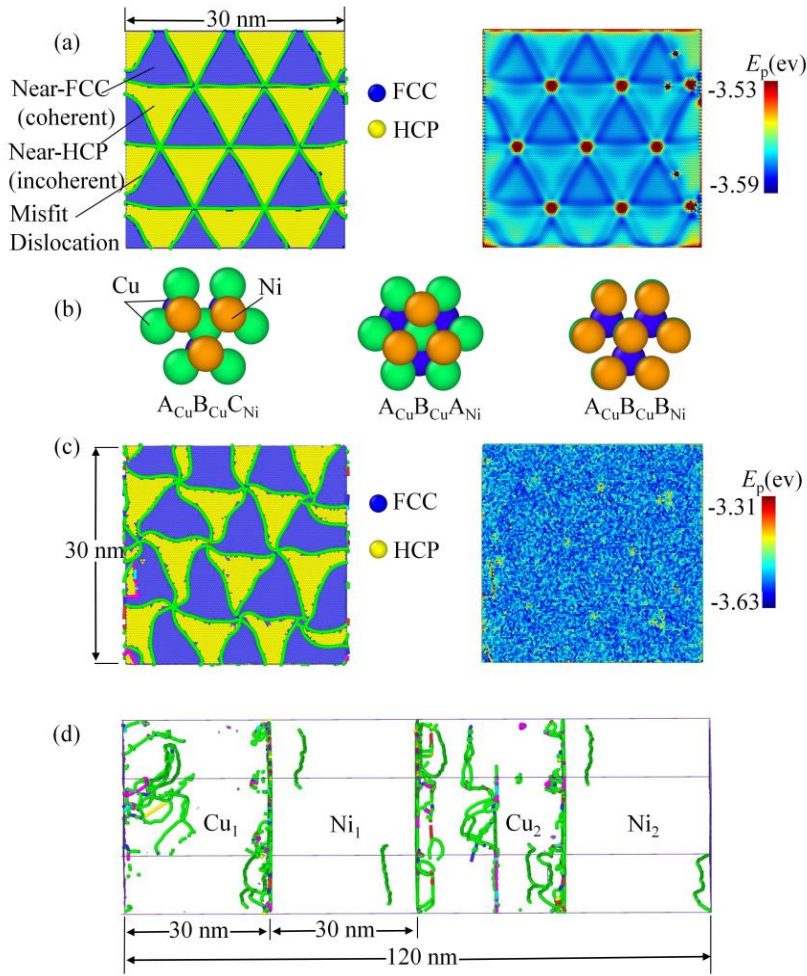

Fig. 2. The initial structure of a Cu(111)/Ni(111) semi-coherent interface in the S1 sample: (a) The misfit dislocations (left) and the potential energy distribution (right) before energy minimization; (b) Illustration of three typical stacking sequences of the Cu(111) and Ni(111) planes adjacent to the interface. Blue and green atoms are Cu; gold atoms are Ni. (c) The misfit dislocations (left) and the potential energy distribution (right) after energy minimization and thermal equilibration. (d) The residual lattice dislocations in Cu and Ni layers. The dimensions of the S1 sample are illustrated in (a), (c), and (d).

The shock responses are simulated by NEMD simulations conducted by the open-source Large-scale Atomic/Molecular Massively Parallel Simulator (LAMMPS) code (Plimpton (1995)). Shockwaves in the target samples are generated by the impact of a Cu flyer in the $x$ ([111]) direction. The initial velocity of the target is zero. The velocity

Verlet algorithm is adopted to solve the MD equations. The time-step in NEMD simulations is set as 0.2 fs to ensure numerical stability. Different impact velocities ($v$ = 0.2, 0.5, 1.0 km/s) are considered. Free boundary conditions are applied along the shock direction ($x$-direction), while periodic boundary conditions are adopted along the transverse $y$ and $z$ directions. From a continuum perspective, the present simulations mimic 1-D strain shock problems. However, the periodic boundary conditions may promote some artificial dislocation activities. For example, dislocations may interact with their periodic images, inducing artificial annihilation. Although such artificial activities may happen, they do not falsify the main results and conclusions of the present work.

The simulation results are analyzed by post-processing programs developed by the authors (Xiang et al. (2017)), with the help of the open-source visualization software OVITO. The details of the analysis techniques are provided in the supplementary material (SM-1) in the file "SupplementMaterial.pdf."

In the following discussions, σ denotes the Cauchy stress tensor. The tensor $p = -σ$ is sometimes used for convenience. Given an interface, the stress tensor in materials at the left and right sides of the interface are denoted as $σ^-$ (or $p^-$) and $σ^+$ (or $p^+$), respectively.

## 3. Characteristics of stress-wave profiles

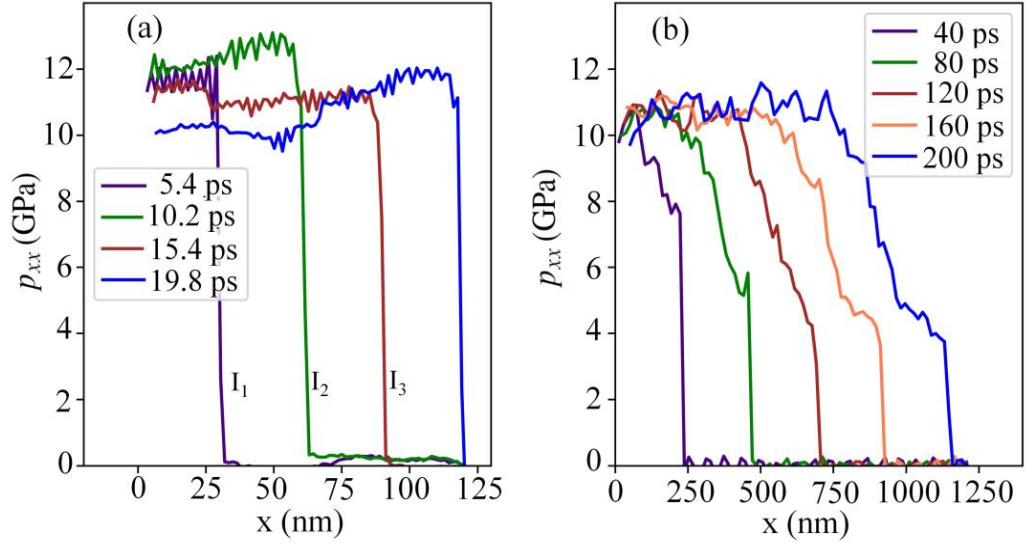

Fig. 3. The wave profiles of $p_{xx}$ at different time instants under $v_L = 0.5$ km/s: (a) in the short S1 sample; (b) in the long S2 sample. In (a), $I_1$, $I_2$, and $I_3$ mark the Lagrangian positions of the three interfaces in S1.

In 1-D strain problems, the $x$, $y$, and $z$ directions of the Cartesian coordinate system are the three principal stress directions; thus, $p_{xy} = p_{xz} = p_{yz} = 0$. Therefore, the stress state of a material point is completely determined by the three normal components $p_{xx}$, $p_{yy}$, and $p_{zz}$. Firstly, the wave profiles of the normal stress component ($p_{xx}$) in the shock direction are discussed. Fig. 3(a) shows the $p_{xx}$ wave profiles in the S1 Cu–Ni sample at different times. Comparison of the wave profiles at $t = 5.4$ ps and $t = 10.2$ ps shows that the transmission stress at the wave front is amplified when the front passes interface $I_1$ from $Cu_1$ to $Ni_1$. This amplification of the wave-front stress ($p_{xx}$) by interfacial reflection is referred to as "reflection-loading." Comparing the wave profiles at $t = 10.2$ ps and $t = 15.4$ ps shows that the wave front stress decreases when the wave front passes interface $I_2$ from $Ni_1$ to $Cu_2$. This decrease in the $p_{xx}$ stress by interfacial reflection is referred to as "reflection-unloading."

We define the transmission coefficient at interfaces as $\eta = p^+_{xx}/p^-_{xx}$. We calculate that $\eta_1 = 1.053 > 1$, $\eta_2 = 0.883 < 1$ and $\eta_1 \times \eta_2 < 1$. The simulations show that reflection-loading occurs when the shockwave propagates from Cu into Ni across the odd-numbered interfaces $I_1$, $I_3$, $I_5$, $\cdots$; reflection-unloading occurs for shockwave propagation from Ni into Cu across the even-numbered interfaces $I_2$, $I_4$, $I_6$, $\cdots$. For any two neighboring interfaces $I_i$ and $I_{i+1}$, reflection-loading occurs at one interface and reflection-unloading at the other. Thus, the transmission coefficients of the interfaces satisfy $(\eta_i - 1)(\eta_{i+1} - 1) < 0$. Moreover, the stress decrement caused by reflection-unloading at one interface is greater than the stress increment caused by reflection-loading at the other interface. This is expressed in the form of transmission coefficients as $\eta_i \eta_{i+1} < 1.0$, $\forall i = 1, 2, 3, \cdots$. This inequality describes the basic unit process for stress attenuation in the Cu–Ni multilayers. It indicates that the wave-front stress decreases after passing through two neighboring interfaces in the multilayers. Based on continuum wave theory, reflection-loading and -unloading at different Cu–Ni interfaces and the resulting stress attenuation correlate with the mismatched Hugoniot curves of Cu and Ni: for a given particle velocity, the Hugoniot pressure of Ni is higher than that of Cu. A detailed explanation is provided in the supplementary material (SM-2) in the file "SupplementMaterial.pdf."

To further study the shockwave attenuation process in the multilayers at larger spatial and temporal scales, we conducted NEMD simulations on sample S2 with extended length ($L_x$ = 2400 nm) in the shock direction. Fig. 3(b) displays the $p_{xx}$ profiles at different times in the shocked S2 sample. The wave-front stress gradually attenuates as the wave propagates through the sample. Accompanying the attenuation of the wave front, the width of the precursor is increased as the wave propagates in the sample. For a sufficiently long sample, the amplitude of the wave-front stress would reach zero and the strong discontinuity at the wave front would vanish. In this way, the initial incident shockwave would finally transform into an isentropic wave.

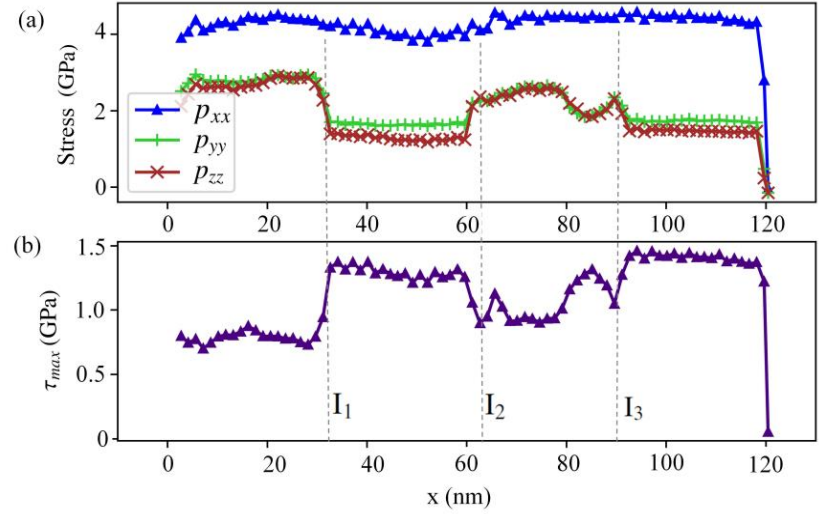

Fig. 4. The profiles of (a) $p_{xx}$, $p_{yy}$, and $p_{zz}$, (b) $\tau_{max}$, and the dislocation distribution in the S1 sample under loading velocity $v_L = 0.2$ km/s at $t = 21.4$ ps, when the shockwave front has just reached the rear free surface.

Although the wave-front stress may increase or decrease upon propagation across interfaces, the stress profile of $p_{xx}$ remains continuous at the interfaces after the wave has passed over the interfaces, as seen in Fig. 3. From a continuum perspective, the continuity of $p_{xx}$ at the interfaces is insured by Newton's third law. The internal traction force acting on the interface due to the stress at the left side is $t^- = \sigma^- \ (1, 0, 0)$, and the internal traction force due to stress at the right side is $t^+ = \sigma^+ \ (-1, 0, 0)$. According to Newton's third law, $t^- = -t^+$, which directly results in $\sigma^- = \sigma^+$ and $p^- = p^+$.

Fig. 4(a) shows the wave profiles of $p_{yy}$ and $p_{zz}$ in sample S1 under loading velocity $v_L = 0.2$ km/s at 21.4 ps, when the shock front has just reached the rear free surface. The profile of $p_{xx}$ is also plotted in Fig. 4(a) for comparison purposes. From Fig. 4(a), $p_{yy} \approx p_{zz} < p_{xx}$. The shear stress is critical for analyzing plastic behaviors. For a given material point, the maximum shear stress over all the planes passing the material point is $\tau_{max} = 1/2(p_{xx} - \min(p_{yy}, p_{zz})) \approx 1/2(\sigma_{xx} - 1/2(p_{yy} + p_{zz}))$ (Fung and Tong (2001)). The profile of $\tau_{max}$ is plotted in Fig. 4(b). The most noteworthy feature in the profiles of the transverse stress components $p_{yy}$ and $p_{zz}$ and the shear stress $\tau_{max}$ is the strong interfacial discontinuities. This differs from the situation in the $p_{xx}$ profile, where continuity is enforced by Newton's third law. The discontinuities in $p_{yy}$, $p_{zz}$, and $\tau_{max}$ are closely related to the mismatches in the elastic–plastic properties of the adjacent layers.

Firstly, elastic property mismatch leads to interfacial discontinuities. In the case of a 1-D strain problem under elastic limits, we obtain:

$$p^-/p^+ = K^-/K^+$$
$$\frac{\tau_{max}^-}{\tau_{max}^+} = \frac{1-K^-}{1+K^+} \tag{1}$$

where $K = C_{12}/C_{11}$ is the elastic constant ratio and $C_{11}$ and $C_{12}$ are elastic constants. The derivation details of Eq. (1) and the calculated elastic constants of Cu and Ni are provided in the supplementary materials (SM-3) in the file "SupplementMaterial.pdf." It is found that $K^{Cu} = 0.365 > 0.289 = K^{Ni}$. As a result, according to Eq. (1), the transverse normal stresses $p_{yy}$ and $p_{zz}$ in a Cu layer should be higher than those in the neighboring Ni layers, and the $\tau_{max}$ in a Cu layer should be lower than that in the neighboring Ni layers. These theoretical predictions are consistent with the direct NEMD simulation results displayed in Fig. 4(a-b).

In addition to the mismatch of elastic properties, plasticity property mismatches also yield discontinuities in the shear stress. This is because the shear stress is limited by the yield strengths of the materials. Based on a single-crystal plasticity model of FCC metals and the Peiers–Nabarro approximation for the critical resolved shear stress $\tau_{crss}$ (Hirth and Lothe (1982)), we can derive the relation of the Hugoniot elastic limits ($p_{HEL}$) at the two sides of an interface as:

$$\frac{p_{HEL}^{Cu}}{p_{HEL}^{Ni}} = \frac{\tau_{crss}^{Cu}}{\tau_{crss}^{Ni}} \times \frac{1-K^{Ni}}{1+K^{Cu}} \approx 0.69 \tag{2}$$

The details of derivation of Eq. (2) are provided in supplementary material (SM-4) in the file "SupplementMaterial.pdf."

According to Eq. (1) and Eq. (2), three situations may occur. Firstly, when $p_{xx} < p_{\text{HEL}}^{\text{Cu}}$, both Cu and Ni layers are in elastic states. In such cases, Eq. (1) holds, and the shear stress discontinuities are induced only by the mismatch in $K$. Secondly, when $p_{\text{HEL}}^{\text{Cu}} < p_{xx} < p_{\text{HEL}}^{\text{Ni}}$, the Cu layer yields and the Ni layer remains elastic. This is the case for the impact loading velocity $v_L = 0.2$ km/s. After the shock, the shear stress in a Cu layer is determined by the yield strength, while that in the neighboring Ni is still determined by the elastic properties. Thirdly, when $p_{\text{HEL}}^{\text{Cu}} < p_{\text{HEL}}^{\text{Ni}} < p_{xx}$, the discontinuity in shear stress results from the mismatch of plastic properties. The simulations with loading velocities $v_L = 0.5$ km/s and 1.0 km/s fall into the third category.

## 4. Interaction of a single lattice dislocation with misfit dislocations

In this section, we investigate the details of the micromechanisms of interaction between lattice dislocations and semi-coherent interfaces. We focus on the interactions between lattice and misfit dislocations, as well as the nucleation of lattice dislocations from semi-coherent Cu–Ni interfaces.

Difficulties in dislocation transmission through interfaces are ascribed to different mechanisms for different interface types, as reviewed by Wang and Misra (2011). For semi-coherent interfaces as in the present work, the resistance to slip is determined by the coherency stresses, Koehler stresses, and misfit dislocations at the interface (Rao and Hazzledine (2000); McKeown et al. (2002); Demkowicz et al. (2008b); Hoagland et al. (2002); Shao and Medyanik (2010); Medyanik and Shao (2009)). The coherency and Koehler stresses can be approximated from the lattice parameters and elastic moduli. The lattice parameter mismatch between Cu and Ni is approximately 2.7%. In order to accommodate this misfit and obtain a coherent Cu–Ni interface, coherency strains (compressive in Cu and tensile in Ni) must be applied to both Cu and Ni layers. The resulting transverse coherency stresses ($\sigma_{yy}$ and $\sigma_{zz}$) were calculated as $p_{coh}$ = 4.54 GPa for {111} coherent Cu–Ni interfaces and 2.38 GPa for {100} coherent Cu–Ni interfaces (Hoagland et al. (2002); Shao and Medyanik (2010); Medyanik and Shao (2009)). The coherent stress is large compared to the yield stresses of the conventional bulk forms of the constituents. In semi-coherent Cu–Ni interfaces, the lattices between the misfit dislocations are coherent. The coherent stresses in the coherent regions remain large, despite the influence of the stress fields of the misfit dislocations. The coherent stresses are important in the interfacial strengthening of Cu–Ni multilayers. The critical stress required to move a dislocation across the interface is approximately equal to the coherency stress (Hoagland et al. (2002, 2004)). The small difference between the coherency stress and the critical stress may arise from contributions from the Koehler stresses and the local stress fields of the misfit dislocations. The Koehler stress is related to the mismatch in modulus (Barnett and Lothe (1974); Rao and Hazzledine (2000)). In Cu–Ni multilayers, the Koehler barrier is ~0.01$\mu$–0.015$\mu$ (Rao and Hazzledine (2000)). As $\mu$ = 54.6 GPa and $\mu$ = 94.7 GPa for Cu and Ni, respectively (Hirth and Lothe (1982)), the Koehler barriers for lattice dislocations in Cu and Ni layers are ~0.55–0.82 GPa and ~0.95–1.33 GPa, respectively.

Misfit dislocations have generally been treated as forest-type obstacles to gliding lattice dislocations in previous studies (Rao and Hazzledine (2000)). The misfit dislocations were thought to offer slip resistance via "cutting" the gliding lattice dislocations (Hoagland et al. (2004)). However, the present simulations reveal new insights on the interactions between lattice dislocations and semi-coherent interfaces.

### *4.1. Local stress-field interaction between lattice and misfit dislocations before contact*

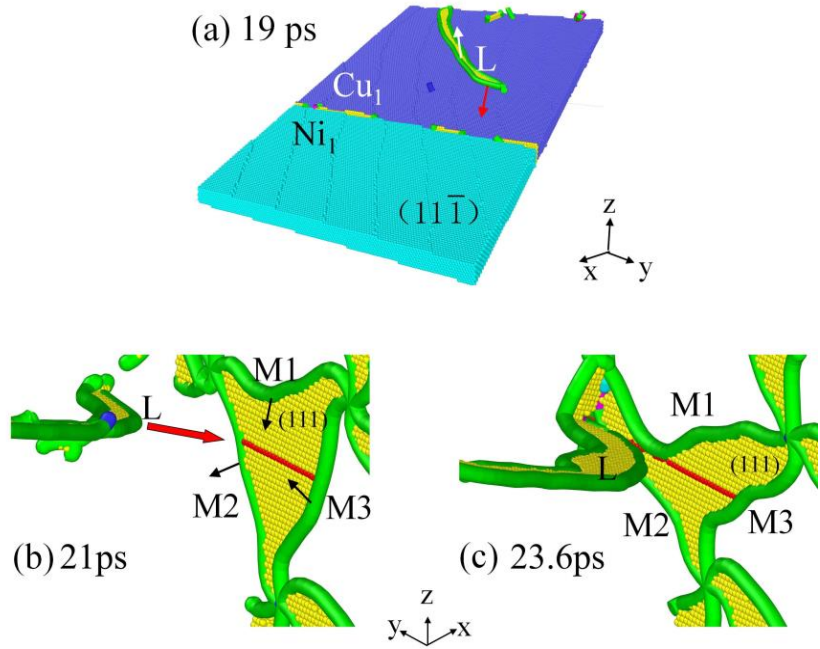

Fig. 5. Motion of the tracked lattice dislocation L and three misfit dislocations M1, M2, and M3 in interface $I_1$, before L contacts the interface. The white arrow in (a) marks the direction of the total Burgers vector of L. The red arrow marks the motion direction of L. Only the HCP atoms that mark the stacking faults are displayed in (b) and (c). The black arrows in (b) mark directions of the Burgers vectors of M1, M2, and M3. The slip plane of L cuts through the stacking fault triangle and intersects with M2 and M3. The red atoms in (b) and (c) mark the intersection line of the slip plane of L and the interface plane $I_1$.

In the case of $u_p$ = 0.2 km/s, the simulation shows that the dislocations are completely trapped in the Cu layers and cannot be transmitted through the semi-coherent Cu–Ni interfaces. In order to obtain more details about the interactions between lattice dislocations and the semi-coherent interfaces, we track the motion of a lattice dislocation (named L) in the S1 Cu–Ni sample. L is an extended dislocation emitted from the impacting surface, slipping on a (1-11) plane in the $Cu_1$ layer towards the first Cu–Ni interface $I_1$, as shown in Fig. 5(a). The extended lattice dislocation L consists of a leading Shockley partial dislocation, a trailing Shockley partial dislocation, and a ribbon of stacking fault between them. The spatial Burgers vectors of the leading and trailing partials are $b_1$ = (−0.6817, −1.2709, 0.2438) Å and $b_2$ =

(−1.3630, 0.0, 0.4876) Å, respectively. The total spatial Burgers vector of L is $b_L = b_1 + b_2 = (-2.04, -1.27, 0.73)$ Å. Finally, the true (local) Burgers vectors of the leading partial and the trailing partial are 1/6[121] and 1/6[211]. Two movies ("supplement-movie-1.gif" and "supplement-movie-2.gif") are provided as supplementary materials for display the slip processes and the reaction of the tracked lattice dislocation L with the misfit dislocations in interface $I_1$.

Fig. 5(b-c) show close-up pictures of the gliding lattice dislocation L and three misfit dislocations, M1, M2, and M3, situated in the interface $I_1$. The three misfit dislocations are bound a triangular misfit stacking fault. The spatial Burgers vectors for the three misfit dislocations are $b_{M1} = (0.0, 0.0, -1.47)$, $b_{M2} = (0.0, 1.2711, -0.7332)$, and $b_{M3} = (0.0, 1.27110.7333)$; the true Burgers vectors are 1/6[1-21], 1/6[112], and 1/6[2-11]. Before the lattice dislocation L reaches interface $I_1$, it interacts with the misfit dislocations through their local stress fields. At $t = 21$ ps, the three misfit dislocations are approximately straight and M1 is nearly parallel to the intersection line, as shown in Fig. 5(b). At $t = 23.6$ ps, the front part of the leading partial of L just contacts interface $I_1$. At this time, the misfit dislocations have all slipped away from their original positions, as shown in Fig. 5(c). Part of M1 touches the intersection line of the slip plane of *L* and the interfacial plane. M2 is obviously twisted. "M3" moves in the direction away from L. The motion of M1, M2, and M3 dramatically changes the shape of the stacking-fault triangle.

MD simulations have revealed the motion of misfit dislocations. However, MD simulations cannot quantify the interaction forces between lattice and misfit dislocations. To investigate the interaction forces, we utilize dislocation theory. In dislocation dynamics calculations, a dislocation line is divided into small segments by a set of discrete nodes. Based on dislocation theory, the force exerted on a given discrete node *K* located at $q_K = (x_K, y_K, z_K)$ by the local stress field of an arbitrary

dislocation line L can be calculated as (Arsenlis et al. (2006)):

$$f_K = \sum_{I \in A_K} \int_{S_{IK}} \frac{s}{\|q_K - q_I\|} \left[ \sigma_{\mathcal{L}}\left(\left(1 - \frac{s}{\|q_K - X_I\|}\right)q_I + q_K\right) \cdot b_{IK} \right] \times \frac{q_K - q_I}{\|q_K - q_I\|} \, ds \quad (3)$$

where $A_K$ denotes the set of discrete nodes connected to node $K$ by a small dislocation segment and $S_{IK}$ denotes the small dislocation segment beginning at node $I$ and terminating at node $K$. $b_{IK}$ is the Burgers vector of $S_{IK}$ and $|q_K - q_I|$ is the length of the segment. $\sigma_L(q)$ is the local stress field of the dislocation L. Based on Eq. (3) and using the dislocation dynamics simulation code ParaDis (Bulatov et al. (2004); Arsenlis et al. (2006)), we calculated the interaction forces between the lattice dislocation L and the misfit dislocations in $I_1$ (including M1, M2, and M3). In our calculations, a constant-strain-rate load type is defined with the strain rate of zero. The mobility law is set to be FCC0-type in ParaDis. As the data of dislocation lines is directly transformed from the MD simulation results, the size of the simulation box in ParaDis calculation is set equal to that of the MD simulation box. The boundary conditions are also set to be consistent with the MD simulations. More specific step-by-step details of the force calculation processes are provided in the supplementary materials (SM-5) in the file "SupplementaryMaterial.pdf."

The forces on the misfit dislocations from the local stress field of L and the forces on L from all the misfit dislocations in interface $I_1$ are plotted in Fig. 6. From Fig. 6(a), L exerts an attractive force on M1 and M2, but a repulsive force on M3. This is consistent with the motion directions of M1, M2, and M3 obtained by direct MD simulations. L exerts a strong attractive force on the segment of M2 near the front of L. Fig. 6(b) shows that the misfit dislocations in interface $I_1$ exert an attractive force on L. Fig. 6(c) shows that the maximum magnitude of the in-plane force component on M1 exerted by L is $\sim 30 \times 10^{-11}$ N. The length of M1 is $\sim 100$ Å and the magnitude of the Burgers vector is $\sim 2.55$ Å. Thus, the maximum effective shear stress on M1 exerted by $L$ is $\tau \approx 30/(2.55 \times 100) \approx 0.12$ GPa. A 1-D shock compression load would not induce shear stress on the interfacial plane. As a result, the local stress field of L is the only factor inducing misfit dislocation slippage in the interfacial plane. From Fig. 6(d), the maximum magnitude of the in-plane force component on L from the misfit dislocations is $\sim 150 \times 10^{-11}$ N. The length of L is $\sim 400$ Å. Thus, the maximum effective shear stress on L due to the misfit dislocations is $\tau \approx 150/(2.55 \times 400) \approx 0.15$ GPa when L is very close to the interface. This stress is very small compared to the Koehler stress barrier ($\sim 0.01$–$0.015\mu = 0.55$–$0.82$ GPa according to Rao and Hazzledine (2000)) and the coherent stress ($\sim 4.54$ GPa, according to Shao and Medyanik (2010)). Thus, the interaction force exerted on L by the misfit dislocations is not the predominant factor for the blocking and transmission of L across the interface. Instead, the reaction of L with the misfit dislocations after it reaches the interface contributes to blocking the motion of L. This will be discussed in detail in the next section.

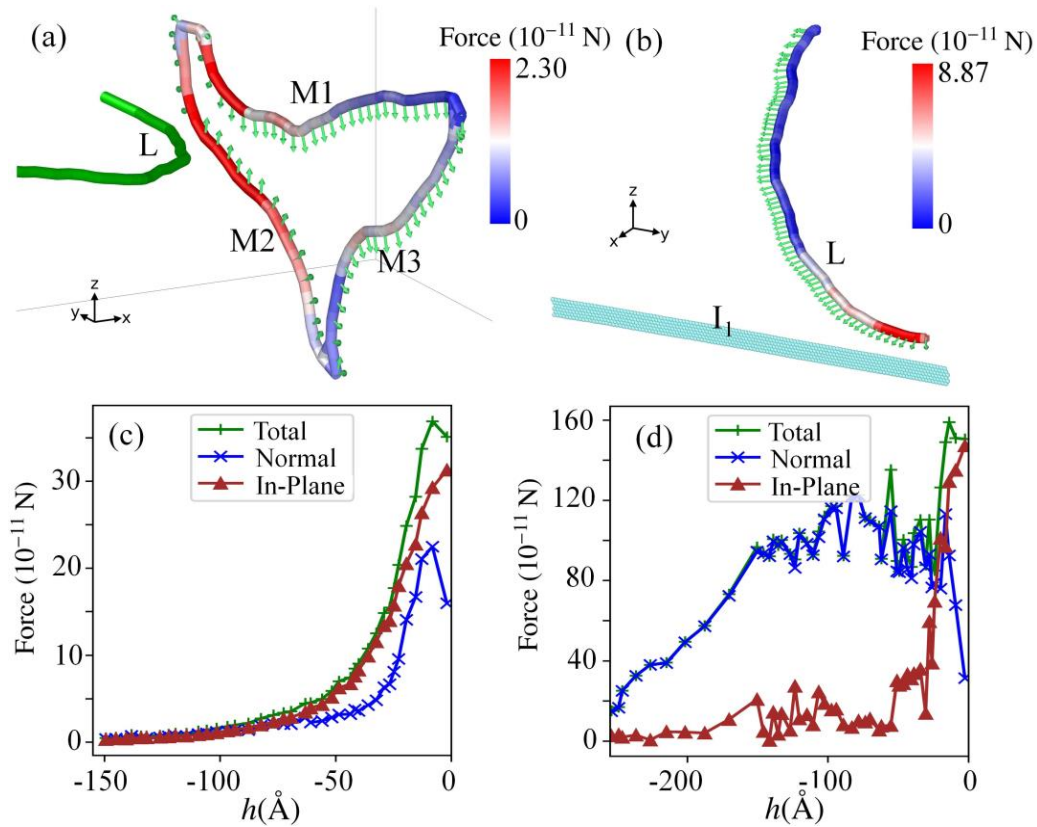

Fig. 6. The interaction forces between L and the misfit dislocations before L contacts the interface $I_1$, calculated by ParaDis. (a) The in-plane components of forces on M1, M2, and M3 exerted by L at $t = 22.6$ ps. (b) The in-plane components of forces on L exerted by all the misfit dislocations in interface $I_1$. (c) The magnitude of the resultant force on M1 due to L, as a function of $h$, where $h = \max_L(x − x_{I_1})$ is the distance of L from the interface. (d) The magnitude of the resultant force on L due to all the misfit dislocations in $I_1$, as a function of $h$.

## 4.2. Reaction between lattice and misfit dislocations after contact: formation and dissociation of hybrid Lomer–Cottrell locks

When the front part of the leading Shockley partial of L reaches interface $I_1$ at $t = 23.6$ ps, it meets the section of M2 attracted by L to the intersection line, as shown in Fig. 7(a). Then the intersecting parts of L and M2 combine to form a stair-rod dislocation S, as shown in Fig. 7(a). Under compression, the leading partial of L continues to attract and combine with M2 at the intersection line; the length of the stair-rod S thus grows along the intersection line. At $t = 31$ ps, the formed stair-rod S has run through the entire misfit stacking-fault triangle, as shown in Fig. 7(b). The reaction between L and M2 can be represented by:

$$\frac{1}{6}[\bar{2}1\bar{1}]\,(\text{"L"}) + \frac{1}{6}[1\bar{2}1](\text{"M2"}) \rightarrow \frac{1}{6}[\bar{1}\bar{1}0](\text{"S"}) \qquad (4)$$

This reaction leads to a change in $b^2$ (the square of the length of the Burgers vector): $a^2/6 + a^2/6 \rightarrow a^2/18$. The combination of L and M2 reduces $b^2$ by $5a^2/18$. According to Frank's law, the energy of a dislocation is proportional to $\mu b^2$. Thus, the reaction of L and M2 to form the stair-rod S is energetically favorable. The stair-rod dislocation S is formed by combining a lattice Shockley partial and a misfit Shockley partial. We refer to this kind of stair-rod dislocation as "hybrid stair-rod dislocations" to distinguish them from the stair-rods in a pure FCC lattice, which are formed by combining two lattice Shockley partial dislocations in different slip planes.

Fig. 5 and Fig. 7 display the motion history of M1. Before L contacts the interface,

M1 is firstly attracted by $L$ toward the intersection line of the slip plane of $L$ and the interface plane $I_1$, as shown in Fig. 5(c). However, after L meets and reacts with M2, M1 is repelled from the intersection line, as shown Fig. 7(a-b). This is because the formed hybrid stair-rod $S$ exerts a repulsive force on M1. In fact, if M2 were absent, the leading partial of L could combine with M1 through:

$$\frac{1}{6}[\bar{2}1\bar{1}] \text{ ("L")} + \frac{1}{6}[\bar{1}\bar{1}2] \text{ ("M1")} \rightarrow \frac{1}{6}[\bar{3}01] \qquad (5)$$

This reaction would cause a change in $b^2$: $a^2/6 + a^2/6 \rightarrow 5a^2/18$. The right-side $b^2$ is lower than that of the left side by $a^2/18$. Thus, the combination between the leading partial of L and M1 might be possible. However, the combination of L with M2 would cause a much greater energy reduction than the combination of L with M1 ($5a^2/18$ vs. $a^2/18$). Thus, it is energetically more favorable for L to combine with M2 than with M1. After the reaction of L and M2, the formed stair-rod $S$ exerts a repulsive force on M1. This is because the combination of $S$ and M1 would require an increase in energy. The stair-rod partial $S$ therefore exerts a repulsive force on both the misfit Shockley partial M1 and the trailing Shockley partial of L. $S$, M1, and the trailing partial of L form a stable, sessile arrangement. In a pure FCC lattice, such an arrangement is referred to as a Lomer–Cottrell lock. In the present situation, the arrangement is somewhat different from that in a pure FCC lattice. Here, one of the Shockley partials is a lattice dislocation, while the other is a misfit dislocation. The stair-rod lies on the intersection line of the slip and interfacial planes. We refer this arrangement as a "hybrid Lomer–Cottrell lock" to distinguish it from the Lomer–Cottrell locks formed in pure FCC lattices. Because the stair-rod partial $S$ is sessile, the lattice dislocation L is trapped at interface $I_1$ by $S$ and cannot be transmitted across $I_1$. The formed hybrid Lomer–Cottrell lock also acts as a barrier to the gliding motion of other dislocations on both the slip plane of $L$ and $I_1$.

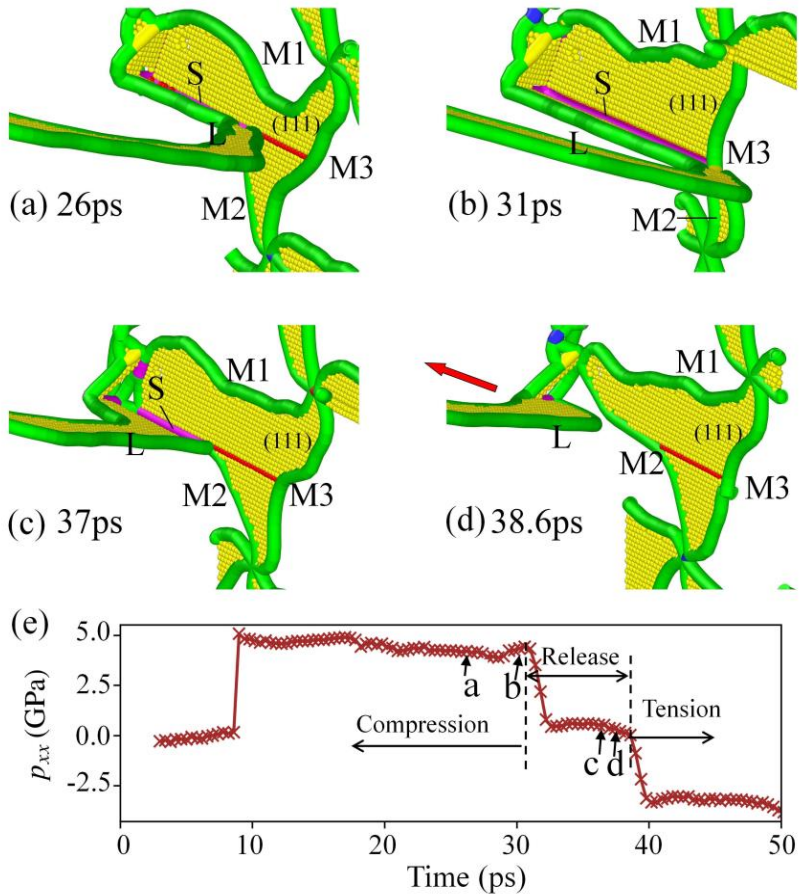

Fig. 7. Reaction of lattice dislocation L with the misfit dislocations, after L contacts interface $I_1$. (a-b) Formation process of stair-rod $S$; (c-d) Dissociation process of $S$; (e) The stress history of interface $I_1$, which is closely related to the formation and dissociation of $S$.

The formed hybrid Lomer–Cottrell lock is not fully stable. From Fig. 7(c-d), the formed hybrid Lomer–Cottrell lock begins to dissociate. At $t = 37$ ps, the hybrid stair-rod $S$ is partly dissociated. At $t = 38.6$ ps, $S$ has completely dissociated. After that, the dissociated lattice dislocation $L$ moves backward into the $Cu_1$ layer.

The dissociation of the hybrid stair-rod (the reverse reaction of Eq. (4)) cannot spontaneously occur because it causes an increase in the dislocation energy. Thus, the dissociation of the hybrid stair-rod must be correlated with external loading. In the case of $v_L = 0.2$ km/s, as displayed in Fig. 7, the dissociation of the hybrid Lomer–Cottrell lock is related to the reverse yielding of materials during the release process. To show this, the time history of the normal stress $p_{xx}$ is plotted in Fig. 7(e). Generally, under shock loading, the time history of $p_{xx}$ for any given material element can be divided into the three stages of compression, release, and tension (Xiang et al. (2013a)). From Fig. 7(e), at $t \approx 31$ ps, the rarefaction wave originating from the reflection of the shockwave at the free surface of the flyer reaches interface $I_1$. As a result, the stress state of $I_1$ turns from compression to release. In the compression stage, Fig. 4(a) shows that the normal stress in the shock direction $p_{xx}$ is higher than the transverse normal stresses $p_{yy}$ and $p_{zz}$, and, therefore, the shear stress $\tau_{max} = 1/2(p_{xx} - 1/2(p_{yy} + p_{zz})) > 0$. During the release stage, the release of $p_{xx}$ is faster than those of the transverse components $p_{yy}$ and $p_{zz}$ (Xiang et al. (2017)). As a result, the shear stress is $\tau_{max}$ first reduced from a positive value to zero. Then, as the release continues, the shear stress becomes negative, as shown in Fig. 7(e). The increase of shear stress in the opposite direction causes yielding again, known as "reverse yielding" in shockwave theory (Yang et al. (2007)). Because the direction of the resolved shear stress is reversed, the reverse yielding induces reverse motion of the dislocations. This is why the dissociated lattice dislocation L moves backward into $Cu_1$ after $t = 38.6$ ps.

The formation and dissociation of hybrid Lomer–Cottrell locks and hybrid stair-rod dislocations are not occasional events; they occur commonly in all the simulations. Fig. 8(a) shows the dislocation configuration at interface $I_1$ in the case of $v_L = 1.0$ km/s, after the lattice dislocations have reacted with the misfit dislocations at $t = 16$ ps. Most misfit Shockley partial dislocations have become hybrid stair-rod dislocations. Fig. 8(b) displays time histories of the total lengths of the stair-rod dislocations in interface $I_1$ under different loading conditions. From Fig. 8(b), as the shock intensity increases, more lattice dislocations react with the misfit dislocations to form hybrid Lomer–Cottrell locks. In all cases, the total length of stair-rod dislocations firstly increases to a saturated value under shock compression. Then some of the hybrid stair-rods dissociate; therefore, the total length of stair-rod dislocations decreases.

To summarize, the simulations reveal that the formation of hybrid stair-rods and hybrid Lomer–Cottrell locks is a major mechanism for blocking the slip motion of lattice dislocations by misfit dislocations. One of the major differences between the present results and previous work is our emphasis on the mobility of misfit dislocations under the local stress field of lattice dislocations. If the misfit dislocations were immobile, they could not move to the intersection line of the slip and interfacial planes. There would therefore be little chance for the gliding lattice dislocation to meet and react with an immobile misfit dislocation to form a sessile Lomer–Cottrell lock. Thus, the previous treatment of misfit dislocations as forest-type obstacles offering slip resistance via cutting, as in Rao and Hazzledine (2000), was reasonable because it did not consider the mobility of misfit dislocations.

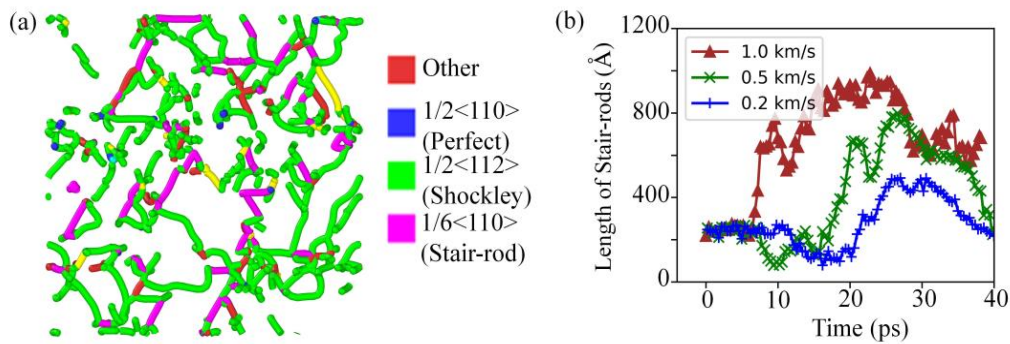

Fig. 8. (a) The dislocation configuration in interface $I_1$ of sample S1 under $v_L = 1.0$ km at $t = 16$ ps, showing the amount of stair-rod dislocations. (b) The total length of stair-rod dislocations in interface $I_1$ of S1 as a function of time during compression and release stages, under different loading intensities.

### *4.3. Slip transmission of a lattice dislocation across semi-coherent interface*

In the case of $v_L = 0.5$ km/s, although the semi-coherent interfaces act as strong barriers for dislocation transmission, some of the lattice dislocations can be transmitted across semi-coherent Cu–Ni interfaces. In this case, we also track the motion of a chosen single lattice dislocation in the S1 Cu–Ni sample. Without introducing any ambiguity, we also label this tracked dislocation L, as shown in Fig. 9(a). As the tracked lattice dislocation L reaches the interface $I_1$, the leading partial of L is cut into several segments by the misfit dislocations. Some segments contact the misfit stacking fault regions, while other segments contact the coherent regions, as shown in Fig. 9(b). Similar to the case of $v_L = 0.2$ km/s, the segments that contact with the misfit stacking fault regions attract and react with some misfit dislocations to form hybrid stair-rods and Lomer–Cottrell locks. At $t = 13.2$ ps, two hybrid stair-rods have formed at interface $I_1$, as shown in Fig. 9(b). The segments that contact with the coherent regions have bowed out into the $Ni_1$ layer. The two ends of the transmitted segments are fixed at the interface by the formed stair-rods. As a result, the transmitted segments attempting to escape from the interface are seriously bent.

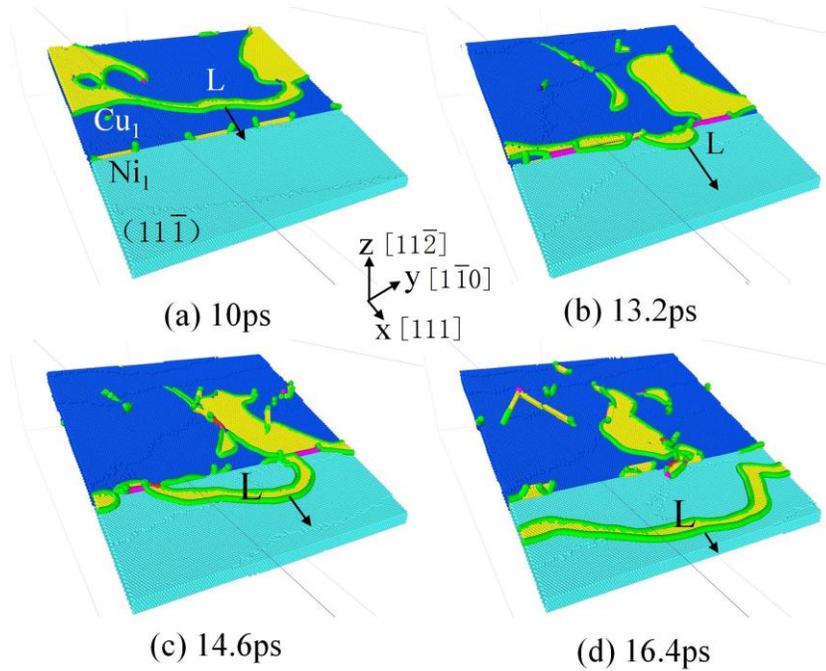

Fig. 9. Slip transmission of the lattice dislocation L across interface $I_1$ (from $Cu_1$ into $Ni_1$), in the S1 sample under $v_L = 0.5$ km/s. (a) Motion in Cu towards the interface; (b) Bowing out in coherent regions and trapping by formation of hybrid stair-rods in incoherent regions; (c) Dissociation of the hybrid stair-rods; (d) Escape from the interface.

The formed hybrid stair-rod gradually dissociates under the following continuous stress loading, as shown in Fig. 9(c). The continuous stress loading leads to dissociation of the hybrid stair-rods. After the lattice dislocation segments are released by the dissociation of the hybrid stair-rods, they continue to move forward into $Ni_1$ and finally escape from interface $I_1$, as displayed in Fig. 9(d). A movie "supplement-movie-3.gif" is provided as an intuitive display of the slip transmission processes.

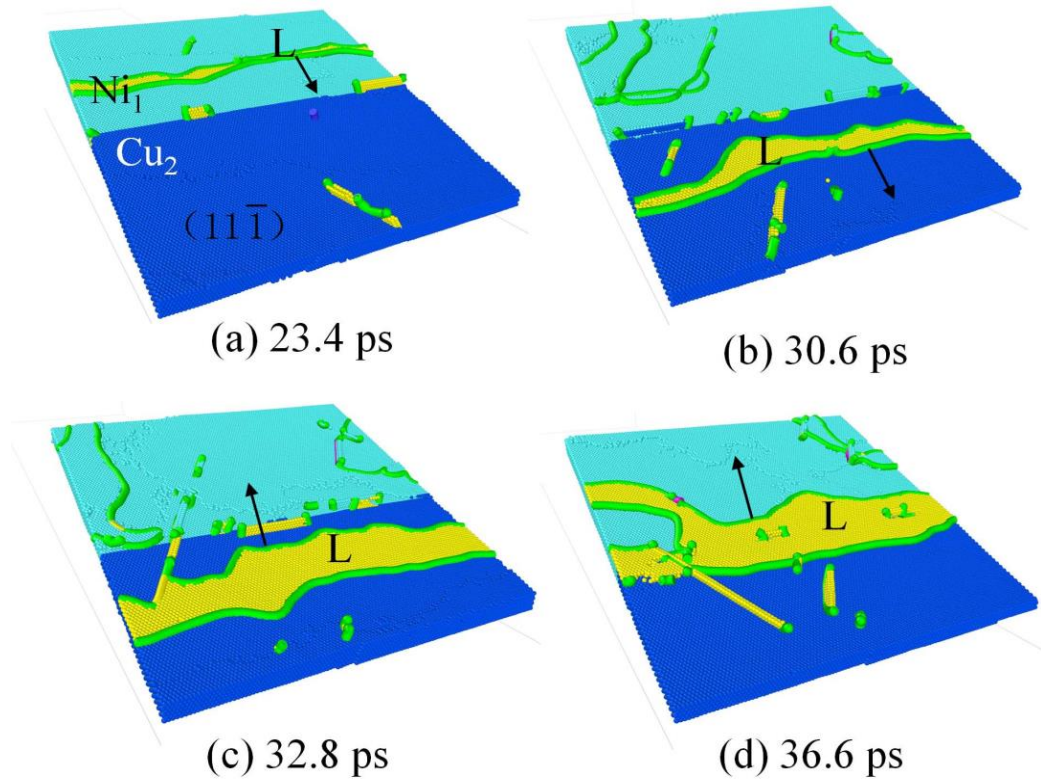

Fig. 10. Slip transmission of the lattice dislocation L through the second semi-coherent Cu–Ni interface $I_2$ (from $Ni_1$ into $Cu_2$), in the S1 sample under $v_L = 0.5$ km/s.

The tracked lattice dislocation L, which has passed across interface $I_1$ from $Cu_1$ into $Ni_1$, can also be transmitted across interface $I_2$ from $Ni_1$ into $Cu_2$. The latter process is displayed in Fig. 10(a-b). We found that the slip transmission processes are direction-dependent. The dislocation $L$ is transmitted much more smoothly across interface $I_2$ than across interface $I_1$. Firstly, almost no hybrid stair-rods are formed at interface $I_2$ during slip transmission of L across $I_2$. Secondly, the dislocation L is not as seriously bent when it passes through interface $I_2$ as when it passes through interface $I_1$. According to dislocation theory, the stress required to bend a dislocation to a radius $R$ is proportional to $\mu b/R$. Higher bending curvatures indicate higher stress barriers offered by the misfit dislocations. Thus, the stress barrier offered by interface $I_2$ is smaller than that offered by interface $I_1$. Moreover, it takes ~5.6 ps for L to pass through $I_1$, while it takes only ~2.6 ps for L to pass through $I_2$. These observations indicate that the slip resistance to L at interface $I_2$ is smaller than at interface $I_1$. In other words, the dislocation is transmitted more easily from a Ni layer to a Cu layer than from a Cu layer to a Ni layer. This direction-dependent feature is attributed to different contributions of the coherent stress at different interfaces. We denote the coherent transverse stress as $p_{coh}$ and the transverse stress components due to the shock compression loading (without the influences of coherent strain) as $p_{yy} \approx p_{zz} = p_0$. Near a semi-coherent interface, the coherent strain generates transverse tension in the Ni layer and transverse compression in the Cu layer, $p_{coh}^{Cu} > 0$ and $p_{coh}^{Ni} < 0$. In the shock compression case, $p_{xx} > p_{yy} \approx p_{zz}$. For simplicity, here we assume that plastic slipping is driven by the shear stress $\tau = 1/2(p_{xx} - (p_{yy} + p_{zz})/2)$. If the coherent stress is absent, then the total shear stress is $\tau_0 = 1/2(p_{xx} - p_0)$. The superposition of the loading and coherent stresses leads to $p_{yy} \approx p_{zz} = p_0 + p_{coh}$, which yields the total shear stress $\tau = 1/2(p_{xx} - (p_0 + p_{coh}))$. Consequently, a tensile coherent stress ($p_{coh} < 0$) increases the total shear stress ($\tau > \tau_0$), while a compressive coherent stress decreases the total shear stress ($\tau < \tau_0$). Applying this result to the present situation, the coherent stress exerts a drag force upon dislocations moving in the Cu layer towards the interface and those moving in the Ni layer attempting to escape from the interface. Thus, the coherent stress retards dislocation motion from Cu into adjacent Ni layers, but facilitates dislocation transmission from Ni into the adjacent Cu layers. Similar insights were presented by Rao and Hazzledine (2000). This explains why dislocation L is transmitted much more smoothly across interface $I_2$ than across $I_1$.

After the L has been transmitted from $Ni_1$ into $Cu_2$, the stacking fault between the leading partial and the trailing partial is expanded. This is because the width of the stacking fault is determined by $d = \mu b^2/4\pi\gamma$, where $\gamma$ is the stacking fault energy and $\gamma_{Cu} < \gamma_{Ni}$. Similar to the case of $v_L = 0.2$ km/s, during the release stage, reverse yielding and reverse motion of L is also observed, as displayed in Fig. 10(c-d). During the release stage, the width of L is largely expanded, which is attributed to the stacking fault energy on release being lower than the stacking fault energy under compression (Zhang et al. (2015)).

### 4.4. Nucleation of lattice dislocations from semi-coherent Cu–Ni interfaces

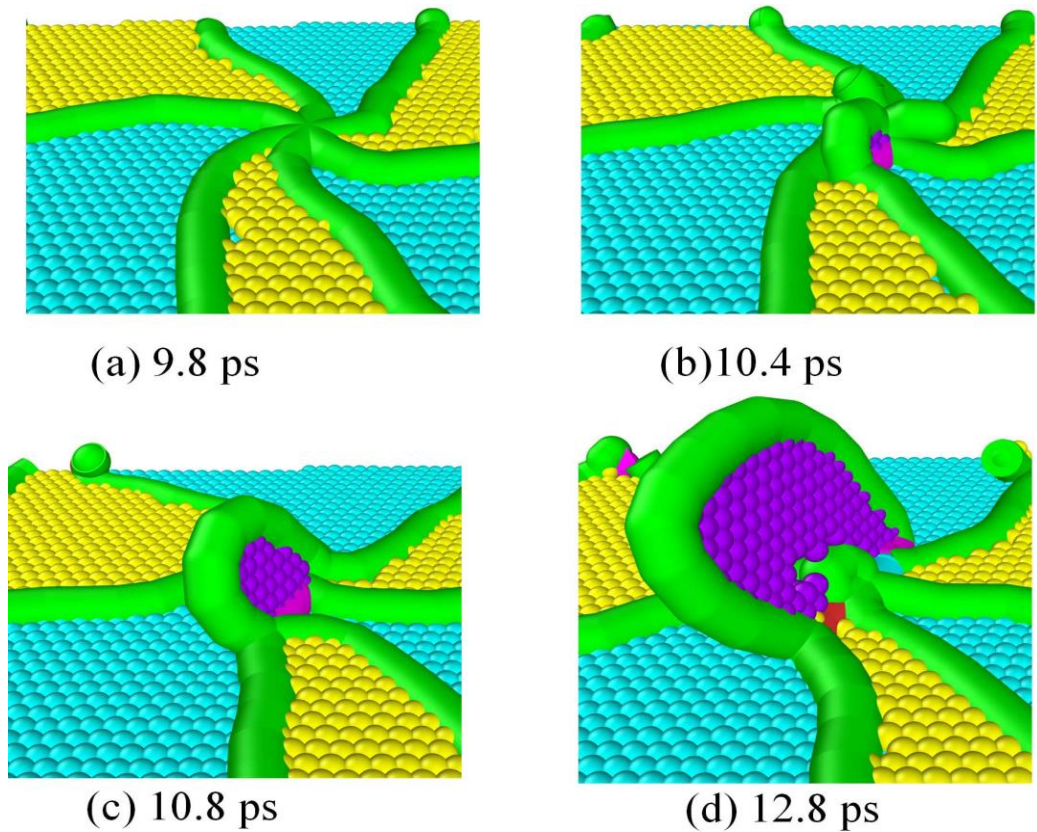

Fig. 11. Nucleation of a lattice dislocation from a misfit node in the semi-coherent interface I$_2$, under $v_L = 1.0$ km/s.

Under relatively weak shock loadings with $v_L = 0.2$ km/s and $v_L = 0.5$ km/s, plastic relaxation occurs mainly through the motion and multiplication of residual dislocations and their interactions with the interfaces. In these cases, dislocation nucleation at the interfaces is not observed. Only when the shock intensity increases to $v_L = 1.0$ km/s is the nucleation of a lattice dislocation from a misfit node observed at interface $I_2$, as displayed in Fig. 11. In a semi-coherent interface, the strain (or stress) is often concentrated at misfit nodes and dislocations. The concentration of strain (or stress) at nodes is higher. As a result, misfit nodes often serve as dislocation nucleation sites.

Despite the various nucleation mechanisms of lattice dislocations, as pointed out by Shao et al. (2015), two fundamental conditions must be satisfied for a nucleation event to occur. Firstly, the interfacial dislocation that serves as the nucleation source or site must align with the slip trace. Secondly, the dot product of the Burgers vectors of the nucleated lattice dislocation and the interfacial dislocation must be greater than zero, thereby relieving the associated strain or stress concentration after the nucleation. The misfit node as displayed in Fig. 11(a) is a spiral-pattern node with six Shockley partial dislocation lines emerging from the node. Although these lines are generally curved, they are approximately straight near the center of the node region. Dislocations around the node can also be viewed as three misfit extended dislocations. Each of the misfit extended dislocations consists of a leading Shockley partial, a trailing Shockley partial, and the intrinsic stacking fault between them. Near the center of the node region, the straight segments of the leading and trailing partials of a misfit extended dislocation combine to form a full dislocation segment with the Burgers vector 1/2<110>. The formed full dislocation segment aligns with the slip trace and thus serves as a nucleation source. With the three pairs of leading and trailing partials, three incipient lattice dislocation segments can arise from the node in the initial stage of nucleation, as shown in Fig. 11(b). The three incipient dislocation segments each belong to different slip planes. However, all three incipient dislocation segments do not ultimately grow to form lattice dislocations. Instead, one of the incipient lattice dislocation segments grows much faster than the other two. This growth significantly decreases the strain (or stress) concentration. The decreased stress concentration decreases the driving force for further growth of the other two incipient dislocation segments. As a result, the other two incipient dislocations are attracted back toward the interface, and only one of the incipient dislocations continues to grow into the Cu layer, as shown in Fig. 11(c). The nucleated lattice dislocation is a leading Shockley loop, followed by a stacking fault. At $t = 12.8$ ps, the trailing Shockley loop nucleates from the interface, as shown in Fig. 11(c). After the trailing Shockley loop is fully nucleated, the whole extended lattice dislocation moves forward into the Cu layer.

## 5. Conclusions

Based on atomistic simulations, dislocation modeling, and continuum elastic–plastic wave theory, we have systematically investigated the shock responses of semi-coherent {111} Cu–Ni multilayers, focusing on the features of dynamical stress-wave evolutions and shock-induced dislocation activities. The main conclusions are summarized as follows:

(1) The influences of the interfaces on the propagating shockwave are elucidated. The simulations reveal reflection-loading (increases in wave-front stress) when the shockwave propagates across an interface from a Cu layer to a Ni layer, and reflection-unloading (decreases in wave-front stress) when the shockwave propagates from a Ni layer to a Cu layer. The reflection-loading and -unloading at different Cu–Ni interfaces are correlated to the mismatch of the Hugoniot curves of Cu and Ni: for a given particle velocity, the Hugoniot pressure of Ni is higher than that of Cu. The calculated Hugoniot curves of Cu and Ni are consistent with the experimental results. When the shockwave propagates over two neighboring interfaces, the stress decrement induced by reflection-unloading at one interface is always greater than the stress increment induced by reflection-loading at the other interface. As a result, the wave-front stress is decreased after it passes through two neighboring interfaces. This is the basic unit process for stress-wave attenuation in multilayers. A simulation on a sample with extended length (~1.2 μm) in the shock direction reveals that wave-front stress attenuation in multilayers is accompanied by widening of the size of the shock precursor. As a result, the incident shockwave would finally transform into a quasi-isentropic wave if the multilayer sample were sufficiently long.

Although normal stress in the shock direction ($p_{xx}$) is continuous, the transverse normal stress components ($p_{yy}$, $p_{zz}$) and shear stresses are strongly discontinuous at the Cu–Ni interfaces. The continuity in $p_{xx}$ is enforced by Newton's third law. The discontinuities in other stress components at interfaces are correlated to mismatches of the elastic–plastic properties in the adjacent layers.

(2) In previous works, interfacial misfit dislocations have generally been treated as forest-type obstacles to gliding lattice dislocations (Rao and Hazzledine (2000)). The misfit dislocations were thought to offer slip resistance via "cutting" the gliding lattice dislocations (Hoagland et al. (2004)). The present simulations provide novel insight into the interactions between lattice and misfit dislocations. The formation of hybrid stair-rods and Lomer–Cottrell locks through the attraction and combination of lattice and misfit dislocations is a major mechanism for interfacial trapping of gliding lattice dislocations. The key aspects of the lattice dislocation–interface interactions are as follows. (*i*) The semi-coherent surface has low shear strength and the misfit dislocations have good mobility. (*ii*) A lattice dislocation approaching the interface generates attractive forces acting on some of the misfit dislocations. (*iii*) The attracted misfit dislocations move towards the intersection line of the slip and interfacial planes. (*iv*) The lattice dislocation intersects and reacts with one of the attracted misfit dislocations to form a hybrid stair-rod. The hybrid stair-rod, a misfit Shockley partial, and a lattice Shockley partial form a hybrid Lomer–Cottrell lock. (*v*) The immobility of the hybrid stair-rod dislocation in the hybrid Lomer–Cottrell lock prevents the transmission of the lattice dislocation across the interface. (*vi*) The formed hybrid Lomer–Cottrell lock blocks dislocation slippage on both the lattice slip plane and the interfacial plane.

(3) Dislocation activities are closely related to the dynamical stress-wave evolution histories. Rarefaction waves, originating from the free-surface reflection of the shock compression wave, may cause reverse yielding of the materials. Reverse yielding induces reverse motion of the dislocations. The formed hybrid stair-rod dislocation is not fully stable, dissociating into a lattice dislocation and a misfit dislocation during the reverse yielding process. The dissociated lattice dislocation moves backward and thus has no opportunity to move across the interface. On the other hand, the formed hybrid stair-rod also dissociates under strong compressive loading. Dissociation of the hybrid Lomer–Cottrell locks facilitates slip transmission of the lattice dislocation.

(4) Slip transmission of a lattice dislocation across a semi-coherent interface is achieved through the following processes: (*i*) A glide lattice dislocation is cut into several segments by the misfit dislocations. (*ii*) The segments contacting the incoherent regions are attracted and trapped by formation of hybrid stair-rods through reaction with the misfit dislocations. (*iii*) The segments contacting the coherent regions are bowed out by overcoming the coherent stress. (*iv*) Under further compression stress loading, the hybrid stair-rods gradually dissociate. (*v*) After a hybrid stair-rod is completely dissociated, the released lattice dislocation escapes from the interface under further stress loading, overcoming the attractive force of the interface.

Slip transmission is direction-dependent. A lattice dislocation is transmitted more smoothly across an interface from Ni to Cu than from Cu to Ni because the role of the coherent stress (or strain) is direction-dependent. The coherent stress (or strain) can either aid or retard the slip transmission of a lattice dislocation. Under shock compression, a tensile coherent stress increases the total shear stress, while a compressive coherent stress decreases the total shear stress. At {111} semi-coherent Cu–Ni interfaces, the coherent stress is tensile in Ni and compressive in Cu. As a result, coherent stress retards dislocation motion from Cu into the adjacent Ni layers. On the contrary, the same coherent stress facilitates dislocation transmission from Ni into the adjacent Cu layers.

(5) Under strong shock loading at the impact velocity of 1.0 km/s, nucleation of lattice dislocations from misfit nodes is observed. Three misfit extended dislocations, and thus three pairs of leading and trailing partials, exist around the misfit node. A pair of leading and trailing partials combines to form a full dislocation segment. The formed full dislocation segment aligns with the slip trace of the lattice dislocations and serves as a nucleation source. The three pairs of leading and trailing partials give rise to three incipient lattice dislocation segments from the node in the initial stage of nucleation. However, only one dislocation has sufficient energy to grow; this ultimately forms a lattice dislocation that slips away from the interface. The other two incipient dislocations are attracted back into the interface by the release of the stress or strain concentration.

The main limitations of the present work include: (*i*) the spall fracture process of multilayers after free-surface reflection of the shockwave is not included in the simulation; (*ii*) the influences of the lattice orientation are not considered. These issues are left for future work; (*iii*) no experimental results are presented. Performing shock experiments is beyond the reach of our research group. We hope that the simulation results in the present work motivate further experimental investigations into the shock responses of Cu–Ni multilayers by using transmission electron microscopy.

## Acknowledgements

We thank Professor Jian Wang of the College of Engineering at the University of Nebraska—Lincoln for valuable suggestions. The work is supported by the National Natural Science Foundation of China (No. 11772068) and the Science Challenging Program of China (No. TZ2016001).